\documentstyle[11pt,aaspp]{article}
\newcommand{\dev}{de Vaucouleurs }
\begin{document}
\title{Two Dimensional Galaxy Image Decomposition}
\author{Yogesh Wadadekar,\altaffilmark{1} Braxton Robbason
\altaffilmark{2} and Ajit Kembhavi \altaffilmark{3}}
\affil{Inter University Centre for Astronomy and Astrophysics,Post Bag
4, Ganeshkhind, Pune 411007, India}
\altaffiltext{1}{yogesh@iucaa.ernet.in}
\altaffiltext{2}{Current Address: Harvard University, robbason@head-cfa.harvard.edu}
\altaffiltext{3}{akk@iucaa.ernet.in}
\begin{abstract}
We propose a two dimensional galaxy fitting algorithm to extract
parameters of the bulge, disk, and a central point source from broad band images of
galaxies. We use a set of realistic galaxy parameters to construct a
large number of model galaxy images which we then use as input to our
galaxy fitting program to test it. We find that our
approach recovers all structural parameters to a fair degree of
accuracy. We elucidate our procedures by extracting parameters for 3 real galaxies -- NGC
661, NGC 1381, and NGC 1427.
\end{abstract}
\section{Introduction}
The luminosity profile of a typical spiral or S0 galaxy most often
contains two components, a spheroidal bulge and a circular disk. If
the galaxy possesses an active nucleus, then a high central point
intensity may also be present. The projected bulge intensity profile
is usually represented by an $r^{1/4}$ law (\dev,~1948), where $r$ is
the distance along the major axis, say.  For  disks 
an exponential law is frequently used (Freeman,~1970). These profiles
are entirely empirical, and have not been derived from a formal
physical theory. However, numerical simulations in simplified
situations  such as those by van~Albada (1982) have been
able to recreate $ r^{1/4} $ profiles for the bulge.

The photometric decomposition of galaxies into bulge and disk and the
extraction of the parameters characterizing these components has been
approached in a number of ways. Early attempts at such decomposition
assumed that the disk would be the dominant component in the outer
regions of galaxies and that the bulge would dominate the inner
regions. 
Disk and bulge
parameters were extracted by fitting for each component separately in
the region in which it was dominant by authors such as Kormendy (1977) and Burnstein (1979).
Schombert and Bothun (1987) employed such a technique for initial
estimation of bulge and disk parameters of simulated galaxy profiles, with
standard laws describing the bulge and the disk with simulated
noise. These parameters were then used as initial input to 
a  $ \chi^2 $ fitting procedure. Tests on simulated profiles
indicated good recovery of both bulge and disk parameters. For a
review of various 1-D decomposition techniques see Simien (1989).
The conventional technique of fitting standard laws to a 1-D intensity
profiles extracted from galaxy images was critically examined by 
Knapen and van der Kruit (1991).
They found that even for the same galaxy different authors derive disk
scale length values with a high average scatter of 23\%. Such large
uncertainty 
in the extracted structural parameters is a hindrance in the study of
structure, formation and evolution of the bulge and disk of galaxies.
Accurate, reliable determination of bulge and disk parameters is a
prerequisite for differentiating between competing galaxy formation
and evolution models.  The conventional one
dimensional technique is also limited because it assumes 
1-D image profiles can be uniquely extracted from galaxy images. This
is not possible if a strong but highly inclined disk is present.

A two dimensional
technique which uses the entire galaxy image  rather than a
major axis profile 
overcomes this difficulty. Such a technique was proposed
by Byun and Freeman (1995), henceforth BF95. A similar approach was
used by R. S. de Jong to extract parameters for a sample of 86 face on
disc dominated galaxies (de Jong, 1996), henceforth DJ96.

In this paper, we describe a 2-D decomposition technique similar to the
one employed in BF95. Extending this work, we  fit for a
central point source in addition to a bulge and a disk.  In addition, our method
takes into account the effects of convolution with a PSF and of photon
shot noise from sky background and the galaxy.  We believe that PSF convolution will
significantly affect parameters, especially those of the bulge. 
We also study the effects of the convolution on the extraction of a
point source.
We also try to  quantify effects of other features such as foreground stars on
parameter extraction. We illustrate the efficacy of our methods by
extracting bulge and disk parameters for three galaxies in two
different filters. We also briefly discuss  reliability of error bars on parameters
extracted.

In the next section, we describe our method of constructing artificial
galaxy images as tests for our bulge--disk--point decomposition procedure. Section 3
describes the decomposition procedure. Section 4 is a detailed
analysis of the testing we performed on it, using artificial galaxy
images..  Section 5 is a description of application of the technique
to three real galaxies. Section 6 contains  a discussion about error
bars on the parameters extracted. Section 7 contains the conclusions.

\section{Simulation of Galaxy Images}

In order to test the bulge and disk decomposition procedure, we have
developed a galaxy image simulation code to generate images closely
resembling those obtained using CCD detectors on optical telescopes. Using the code it is
possible to simulate a CCD image of a galaxy with an arbitrary bulge,
disk and point at an arbitrary position and orientation on the
CCD. The image can be convolved with a given circular or elliptic PSF
and Poisson noise can be added if required. Stars can be introduced into the
image at random positions and additional features of galaxies such as
absorbing dust lanes can be added.  All parameters used by the program
can be easily modified by the user through a parameter file.

Galaxy profiles are the projections of three dimensional luminosity
profiles onto the plane of the sky. The disk is inherently circular,
so it projects as an ellipse. The inclination angle of the disk with
respect to the plane of the sky completely determines the ellipticity
of the disk in the image.  Bulges, taken to be triaxial ellipsoids in
the general case, also project as ellipses. The ellipticity of the
bulge does not reach such high values as the disk. For a triaxial
ellipsoid with major axis $a$, minor axis $b$, and an intermediate
axis $c$, the highest possible ellipticity will be $1-b/c$.  Therefore
the projected galaxy shows elliptical bulge isophotes and in many
cases -- more elliptical disk isophotes.

For an arbitrary point on the  image plane, intensity contribution 
comes from the bulge and the disk, and depends on their respective central
intensities, ellipticities and scale lengths. Near the galaxy center
there is an additional contribution from the 
point source, if one is present.

In our galaxy simulation, the projected bulge component is represented by
the $ r^{1/4}$ \dev law with effective (half light) radius $r_e$ (which is the
radius within which half the total light of the galaxy is contained), intensity at the half
light radius $
I_e $ and a constant ellipticity $ e_b = 1 - \rm (minor\ axis\ 
length/ major\ axis\ length) $:
\begin{eqnarray}
I_{bulge}(x,y) &=& I_e e^{ -7.67 [(r_{bulge}/r_e)^{1/4} - 1]}, \\
r_{bulge} &=& \sqrt{x^2 + y^2/(1 - e_b)^2}, \nonumber
\end{eqnarray}  
where $x$ and $y$ are the distances from the center along the major 
and minor axis respectively. 

The projected disk is represented by an exponential distribution
with central intensity $ I_s $, scale length $ r_s $ and a constant
ellipticity $ e_d $, 
\begin{eqnarray}
I_{disk}(x,y) &=& I_s e^{- r_{disk}/r_s},\\
r_{disk} &=& \sqrt{x^2 + y^2/(1 - e_d)^2}. \nonumber
\end{eqnarray}
The disk is inherently circular. The ellipticity of the disk in the
image is due to projection effects alone and is given by:
\begin{eqnarray}
e_d &=& 1 - \cos(i)
\end{eqnarray}
where $i$ is the angle of inclination of the circular disk with the
image plane.
Finally, the point source is represented by  an intensity added  to the
central pixel of the galaxy prior to convolution.

We use these empirical laws for the bulge and the disk keeping in mind
that they do not satisfactorily describe all galaxies. Features such
as stars can be added if required. Stars are added at random positions
as intensities at a single pixel prior to convolution with the
PSF. The convolution with the PSF is performed in the Fourier
domain. For adding photon shot noise, a constant sky background is
added to every pixel. The resultant count in each pixel (which
includes intensity from the galaxy as well as the sky background) is
multiplied by the gain (electrons/ADU).  A random Poisson deviate
about this value is obtained. The deviate is then divided by the gain
and the background is subtracted out.

The program takes about 15 seconds to generate a galaxy image, when
running  on a Sparcstation 10 for a
256 $ \times $ 256 pixel image. A copy of this code (written entirely
in ANSI C) is available upon request.
 
\section{ 2-D Decomposition Technique}

\subsection{Building The Model}

The model to be fit is constructed using the same procedure as the
simulated images described in the previous section, except that features of the image that are not contributed by the galaxy such as  stars and Poisson noise are not added.

\subsection{The Decomposition Procedure}

To effect the decomposition, we attempt to iteratively minimize the
difference between our model and the observed galaxy (or a simulated one), as measured by
a $\chi^2$ value.  For each pixel the observed galaxy counts are
compared with those predicted by the test model. Each pixel is
weighted with the  variance  of its associated
intensity as determined by photon shot noise  of the combined sky and
galaxy counts at that pixel. Photon shot noise obeys  Poisson
statistics and so the variance is equal to the intensity value. Hence: 

\begin{equation}
\chi_{\nu}^{\,2} = {1\over \nu} \sum_{i,j} {[I_{model}(i,j)-I_{obs}(i,j)]^2 \over I_{obs}(i,j)},
\end{equation} 
where $i,j$ range over the whole image, $\nu = N - $ (number of fitted
parameters)  is the number of degrees of freedom with $N$ being the
number of pixels in the image involved in the fit, and $I_{obs}$ is assumed to be greater than zero
in all cases.

CCD images of galaxies contain features such as foreground stars and
bad pixels that may contaminate the decomposition procedure. We take
care to block out bad pixels and
visible foreground stars before commencing decomposition.  

For real galaxies, our decomposition program assumes that pixel values
$I$ represent real photon counts.  If the image has been
renormalized in any way (divided by the exposure time for example),
the extracted $\chi^2$ value should be multiplied by the appropriate
factor to account for that normalization.


Seven  fitting parameters are used in our scheme. These are $ I_e, I_s,
r_e, r_s,$ $ e_b, e_d,$ and a central point source intensity $I_p$. The first 4 parameters were
used by Schombert and Bothun (1987), and the first six were used by the
authors of BF95 and DJ96.  We have added capability in our code to fit
for position angle and a constant background. During our
preprocessing we estimate the background carefully and subtract it
out. Where it is possible in real cases, fitting for sky background
is  not required. 

The minimization uses MINUIT 94.1, a multidimensional minimization
package from CERN, written in standard FORTRAN 77. MINUIT allows the
user to set the initial value, the resolution, and the upper and lower
limits of any parameter in the function to be minimized. Values of one
or more parameters can be kept fixed during a run. MINUIT has several strategies to perform the
minimization. Our method of choice is  MIGRAD, a stable
variation of the Davidon--Fletcher--Powell variable metric algorithm.
It calls the user function (in our case $\chi_{\nu}^{\,2} $)
iteratively, adjusting the parameters until certain criteria for a
minimum are met.  Our code typically takes 0.1 seconds per iteration on
a SGI Indigo 2 workstation, when working on a 64 $\times$ 64 pixel
galaxy image. On a slower Sparcstation 10, time taken increases by a
factor of about 1.5.  Since we are using fast Fourier transforms to
convolve the model with a gaussian PSF, the execution time can be
expected to scale as $N \log N$, with $N$ the total number of pixels in
the image. We do find that execution time scales almost linearly with
the number of pixels in the image. About 1000 iterations are required
for convergence criteria to be satisfied when all 7 parameters are
kept free, so a typical run takes less than two minutes.  Computation
time is reduced as the number of free parameters is reduced. 

A copy of this code (written mostly in ANSI C with some optional display
routines in IDL) is available
upon request. 

\section{Reliability of Galaxy Image Decomposition}
We conducted  elaborate  tests of the effectiveness of
the program in extracting parameters under different input conditions.
The tests of error in PSF measurements serve as warnings against
claiming too much accuracy in the extraction of a point without very
accurate measurements of the PSF.

\subsection{Large-Scale Testing under Idealized Conditions}

To test the accuracy and reliability of our algorithm, we constructed
a wide range of model galaxies with idealized bulge and disk
components. Images of 500 model galaxies were generated by random 
 uniform selection of intensities, scale lengths and ellipticities. The ranges
we used for each of the parameters are, 

\begin{enumerate}
\item{for the bulge}
\begin{eqnarray*}
18<&\mu_{\em eff}&<24 \ {\rm mag\  arcsec^{-2}},\\
3<&r_e       &< 30 \ {\rm pixels},\\
0.0 <&  e_b&< 0.4,
\end{eqnarray*}
\item{and for the disk}
\begin{eqnarray*}
15.5 <  &\mu_s&  < 21.5 \ {\rm mag\  arcsec^{-2}},\\
3 < &r_s& < 30 \ {\rm pixels}, \\
0 < &e_d&  < 0.9.
\end{eqnarray*}
\end{enumerate}
The ranges chosen for the parameters encompass most values
encountered in real galaxies, and the intensities were chosen to
correspond to the above magnitudes 
using photometric data on a 1m 
class telescope. Note that we fit in intensity space, not magnitude
space, so the values were distributed approximately uniformly in intensity
space. The galaxies generated by the galaxy simulation program
were  then used as input to the bulge disk decomposition program. We
studied how accurate and reliable the decomposition program is in
recovering the input parameters.

We place no additional relative 
constraints on permissible parameter values such as those used in
BF95.  We find that the  $ \chi_\nu^{\,2} $ for the fits is worse
than 2 in only 25 cases out of 500, giving a 5\% failure rate. These failures
are caused by the presence in parameter space of local minima
close to the starting values given to MINUIT, or by one or more
parameters hitting their limits, causing the gradient-driven
minimization routine to fail. It is possible to completely eliminate
such 
failure by changing the initial value and constraining the range of
parameters around the {\em expected} value. Such a procedure cannot be
adopted while working with real
galaxies as it is not possible to constrain the parameters {\em a
priori}. We find that a $\chi_\nu^{\,2} < 2 $ almost always corresponds to
good recovery of input parameters, and a $ \chi_\nu^{\,2} > 2 $ always
corresponds to poor recovery of input parameters. Poor recovery of one
parameter almost always implies poor recovery of all other parameters,
and a high value of  $\chi_\nu^{\,2} $. In the instances of poor
recovery deviations from true values
for different parameters are not correlated, which is consistent with
the high values of $\chi_\nu^{\,2} $ obtained for them. 
\begin{figure}[htb]
\begin{center}
\leavevmode
\epsfbox{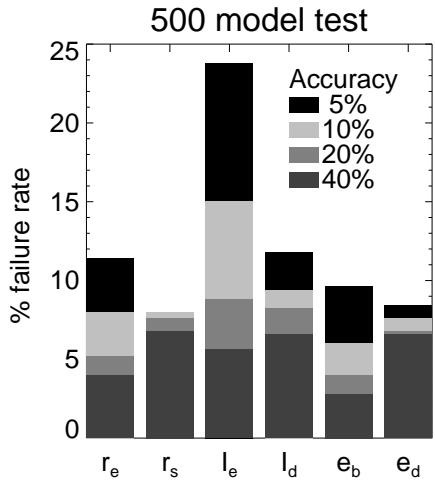}
\caption{Failure level in  500 point test. The Y axis shows percentage
of cases in which the program fails to recover the parameter indicated
on the X axis at a level indicated by the shading (see text)}
\label{fig:500bargraph}
\end{center}
\end{figure}

The parameter errors are summarized in Fig.~\ref{fig:500bargraph}. 
There is a column for each parameter, with the shading indicating which
failure  level the bar represents.  For example, for 5\% of
models, the program failed to extract the bulge scale length with
an accuracy of better than 20\%, and for 7\% it failed to extract the
same parameter with an accuracy of better then 10\%.  Accuracy is
defined here as the the difference of the input parameter and the
recovered parameter, divided by the input parameter and expressed as a
percentage.

\begin{figure}[p]
\epsfbox{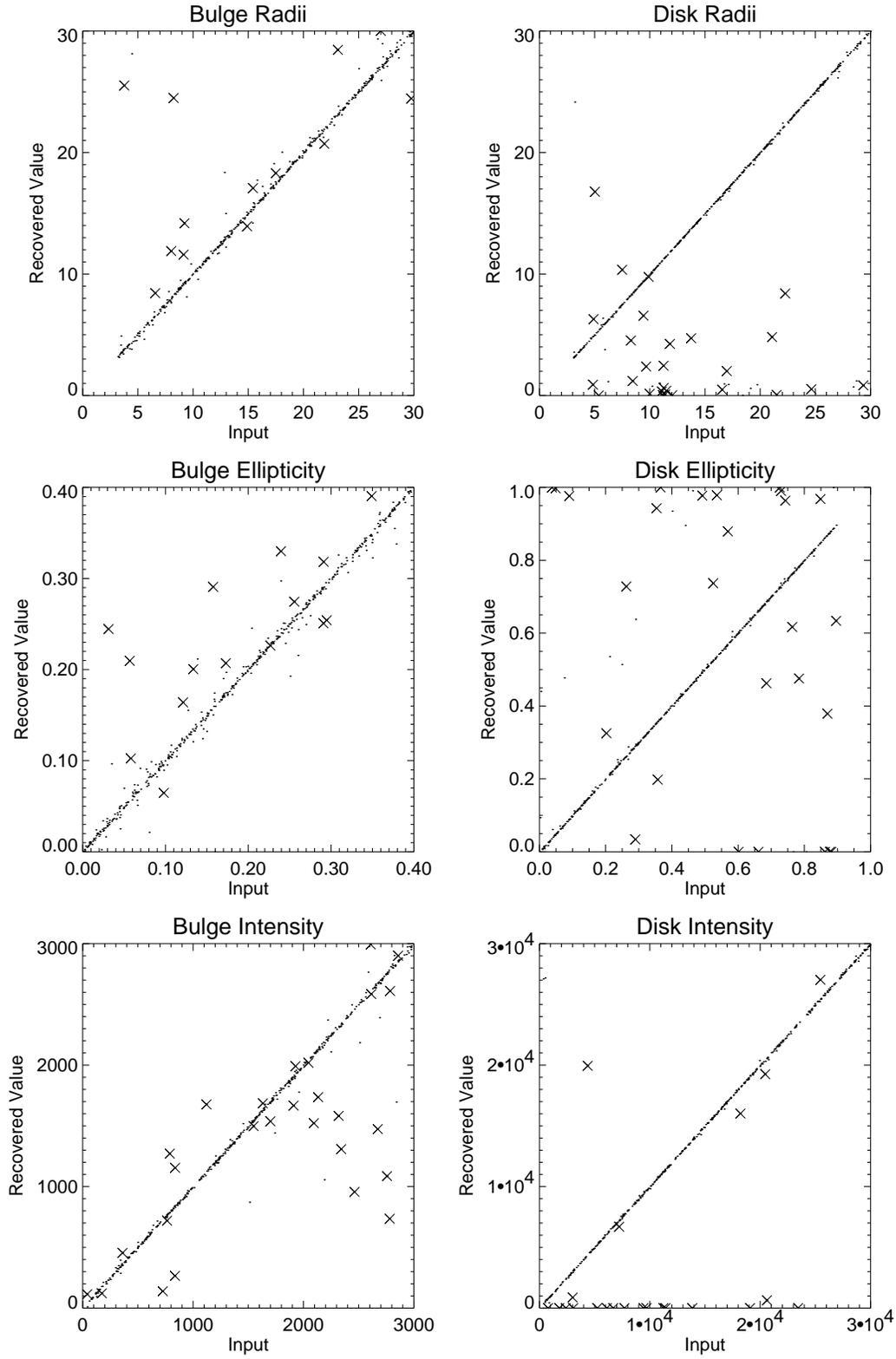}
\caption{Scatter plots of extracted versus input parameters. Parameters
extracted with $\chi_\nu^{\,2} > 2$   are indicated by a $\times$}
\label{fig:500tst}
\end{figure}

From Fig.~\ref{fig:500bargraph} we see that all input parameters except bulge
intensity are extracted 
with less than 10\% error in over 90\% of the test cases. The bulge intensity seems to have been
extracted less accurately  than the other quantities. 
The extracted versus input
data have been plotted 
in Fig. ~\ref{fig:500tst}, with points having  $\chi_\nu^{\,2} > 2$
marked with a $\times$. Note that the points with a bad fit generally
lie far away from the line on which the input value equals the
extracted value. Most of the remaining  points (those with $ \chi_\nu^{\,2} < 2
$) plotted on these graphs are
tightly grouped along this line. There is no systematic trend in the deviation
of extracted parameters from their true value for any of the
parameters plotted. The scatter plot  in
Fig.~\ref{fig:500tst} is tighter than a similar plot presented in BF95.

In a real situation, one would not be running blind,
and if minimization failed with one set of parameters, one would use different
starting values and try again.  
It should be noted that failure in recovering parameters is easy to
detect as it is always flagged   by a high
$\chi^2$ value. We conclude 
that for our test cases at least,  the input parameters are recovered very well, and without
systematic or significant deviations.

\subsection{Tests of stability}
We conducted a series of tests to determine how the program responds
to deviations from the idealized conditions assumed in the previous
section. We looked at some of the  problems encountered when dealing
with real images rather than simulated ones. 

\subsubsection{Effect of changing S/N ratio}
The S/N ratio improves with increase in exposure time. We examined the
image of the same simulated galaxy using pixel counts
for a bright galaxy and sky background corresponding to exposure times ranging
from about 5 seconds to 8 minutes on a 1 \, m class telescope. The
exposure times (and hence the pixel counts) varied by a factor of 96
and S/N ratio by a factor of about 10 ($ \simeq \sqrt{96}$). The background counts used
were estimated from 
observations made on a 1m class telescope in the Cousins R filter. We expected
that as S/N got  better, the fit would improve and parameter recovery
would get more accurate. We found  that the accuracy of the extracted
parameter values  is strongly dependent on the exposure time only for
very short exposures ($<$ 30 seconds) (Table~\ref{tab:sbyn}). Peak
counts of less than one thousand for galaxies are not very useful for
the purpose of bulge disk decomposition.

\begin{table}[h]
\begin{center}
\begin{tabular}{c c c c c c }

\hline
Exp. time & $\chi_\nu^{\,2}$ & Peak & Input  	 & Extracted  &
\% Error in \\ 
 (sec) & & Counts & $I_e$& $I_e$ & $I_e$ \\ \hline
5     & 1.00 & 302 & 4.17 & 6.4 & 53.0\\
15    & 1.00 & 914 & 12.5 & 11.1 & 8.9\\
30    & 1.02 & 1923 & 25    & 24.2 & 3.2\\
60    & 1.04 &	3665 & 50    &	51.2 & 2.4 \\
120   & 1.07 &	7623 & 100   &	101  &1.0 \\
240   & 1.11 &	14826 & 200   &	202  &1.0 \\
480   &	1.14 &	29740 & 400   &	415  &3.75 \\ \hline

\label{tab:sbyn}
\end{tabular}
\end{center}
\caption{Effect of changing exposure time on extraction of bulge intensity}
\end{table}

It is seen that  $\chi_\nu^{\,2}$  {\em increases} slowly but monotonically with exposure time.  This is an
artifact of the way sky background is used  in the program creating
the input galaxies.  When simulating galaxies, background is added,
Poisson noise is calculated using the intensity of both background and
galaxy,
and the background is subtracted out.  Then, when the fitting program
runs, it estimates the noise at each pixel as the square root of the number of
counts at that pixel, but the actual noise is  the square root of the
sum of the number of counts  and the background. This causes the
points with low counts to be weighted more than they should be,
(resulting in higher $\chi_\nu^{\,2}$) but
the difference is small. 

\subsubsection{Effect of erroneous measurement of PSF}

With real data it is often impossible, even if a large number of stars
are used, to measure the PSF to an accuracy of better than about
5\%. One reason is variation of the PSF in different regions of the
CCD. Therefore, it is important to know how the fit will react to an
over-- or under--estimation of the PSF, and to an elliptical PSF.  The
value of the point source is expected to be affected the most because
of errors in PSF estimation. If the bulge scale length $r_e$ is very
small, then bulge parameters will also be seriously affected by an
incorrect estimation of the PSF. We ran two separate tests, one with a
circular PSF, overestimated or underestimated by up to 20\%, and
another where only one axis of the PSF changed by 20\% while the other
remained constant thereby generating elliptical PSFs. For the fitting
we used  a circular PSF with FWHM of 1 pixel  in all the test cases.

\begin{figure}[p]
\plottwo{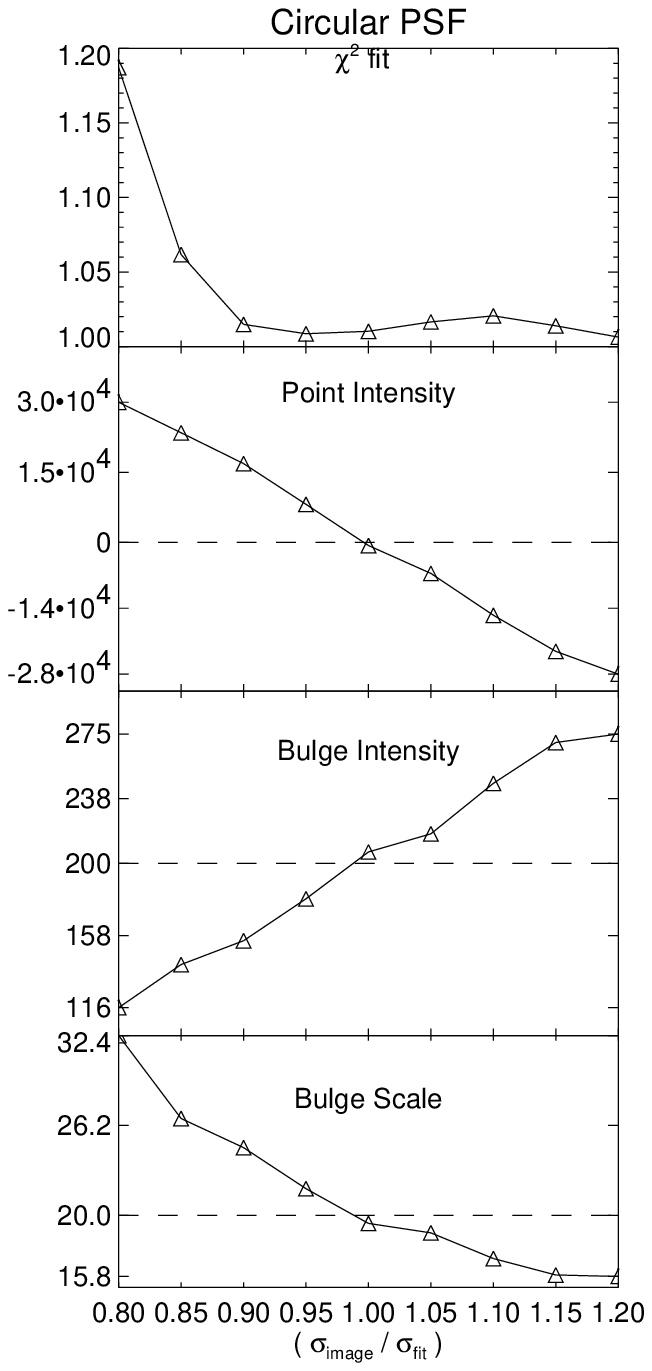}{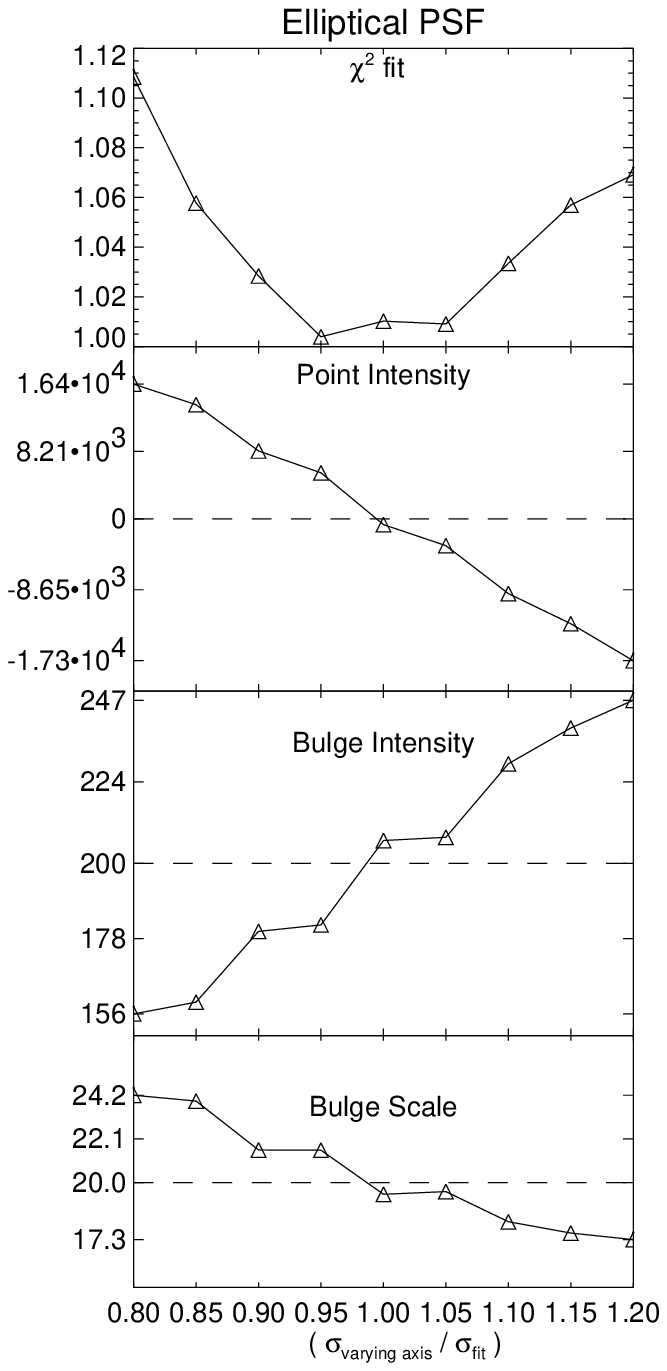}
\caption{Effects of incorrect estimation of PSF's on extracted
parameters. The panels show changes in point intensity, bulge intensity
and bulge scale length under a changing circular PSF (left panel) and
a changing ellipticity of the PSF . The dashed lines indicate  the values of
input intensities, which were held constant. The Y--axis shows the
corresponding fitted intensity. See the text for an explanation}
\label{fig:PSFerr}
\end{figure}

Our simulated  images are generated using a PSF with FWHM $\sigma_{image}$; this 
corresponds to the psf determined by seeing conditions in an actual observing situation.  
The psf used as input to the decomposition program has FWHM $\sigma_{fit}$.  
$\sigma_{image}/ \sigma_{fit}\neq1$ corresponds to the situation where an error is made in
the estimation of $\sigma_{image}$ in an observing run.  Here we are assuming that the 
psf is circular and the only error is in estimating its FWHM.   
When $\sigma_{image}/ \sigma_{fit} < 1$ the spreading of the image due to seeing is 
over estimated.  The excess deconvolved intensity
at the center seen by the decomposition program then causes it to generate
a fictitious point source.  The minimum value for the ratio
we have used in the test is 0.8. At this ratio  the  value of the point is very high  
intensity of the fictitious  point source  is very high, as can be
seen from Fig.~\ref{fig:PSFerr}. The bulge intensity is at its minimum
value.
The intensity of the fictitious point source decreases  and that of
the bulge increases continuously
as $\sigma_{image}/ \sigma_{fit}  \rightarrow 1$
If the FWHM of the psf is underestimated, i.e. $\sigma_{image}/ \sigma_{fit} > 1$,
the point intensity becomes negative\footnote{A negative point is non physical,
of course, but in general we will allow for it because  \dev law
does not hold near the center of most galaxies. In our simulation
however we assume that the law holds right to the center.}.

The variation $\sigma_{image}/ \sigma_{fit}$  did not affect the extracted disk scale length,
which only once  deviated by more than 1 
pixel.  The disk intensity increased with increasing input PSF,
analogous to  the increase in bulge intensity discussed above.
The extracted bulge and point source intensities, and bulge radius, all vary
systematically and approximately linearly with the error in  PSF estimation. 
$\chi_\nu^{\,2}$,  is very
good in all cases, decreasing somewhat as the PSF gets to be closer to our
estimate of 1.  These
results are plotted on the left panel of Fig.~\ref{fig:PSFerr}.

To see the effect of errors in determining the shape of the PSF, we
generated galaxies with different elliptical PSFs. Such PSFs are observed, for example, 
if the plane of the CCD is inclined to the focal plane of the
telescope.  The decomposition program continued to use a fixed circular
PSF. The sequence of image PSFs  was
generated by keeping one of the principal axis of the ellipses always equal to 
$\sigma_{fit}$, and varying the other principal axis so that ratio of the two 
changed from 0.8 to 1.2.  The 
results of parameter extraction are plotted on the right panel of Fig.~\ref{fig:PSFerr},
as a function of $\sigma_{image}/ \sigma_{fit}$, where the ratio is now taken along the
changing principal axis.   

When the PSFs used in the simulation as well as fitting are both circular, 
but unequal, the fitting procedure leads to a positive or negative fictitious point source.
A good overall fit is obtained with $\chi_\nu^{\,2}$  close
to unity in the latter case, i.e.  when $\sigma_{image}/ \sigma_{fit} > 1$, because here
the overall intensity at the centre remains small and  best fit bulge parameters  which
give a good fit, together with the negative point source can be found. In the case
of a positive point source, changed bulge parameters cannot compensate for the error in
the PSF and the quality of the fit is diminished.  In the case of the elliptical
PSFs, $\chi_{\nu}^{\, 2}$ is greater than unity on both sides of $\sigma_{image}/ \sigma_{fit}=1$.

Bulge ellipticity, which was set to 0.1 in all simulated images, 
was extracted very well in the case of the circular PSF as it varied over its range of
FWHM.  When the PSF becomes elliptical, we expect the extracted ellipticity to
increase as well, and it does, but only to 0.12 for the most
elliptic PSF, which had ellipticity 0.2. The ellipticity close to the centre of the
galaxy is of course wholly determined by the shape of the PSF, while
further away, the effects on the extracted ellipticity are much smaller. 

\subsubsection{Fitting in the presence of stars}
We added upto 20 randomly positioned stars to a 128 $\times$ 128 pixel
image, where the star
brightnesses were clearly visible in a magnitude plot, and ran the
program without blocking out the stars.  The presence of the stars
worsened the $\chi^2$ considerably but the extracted parameters were
not affected in any significant or systematic way. Masking out the
stars improved the $ \chi^2 $ to normal values ($\approx 1$), without
affecting the accuracy of the parameter recovery.

\begin{figure}[h]
\epsfbox{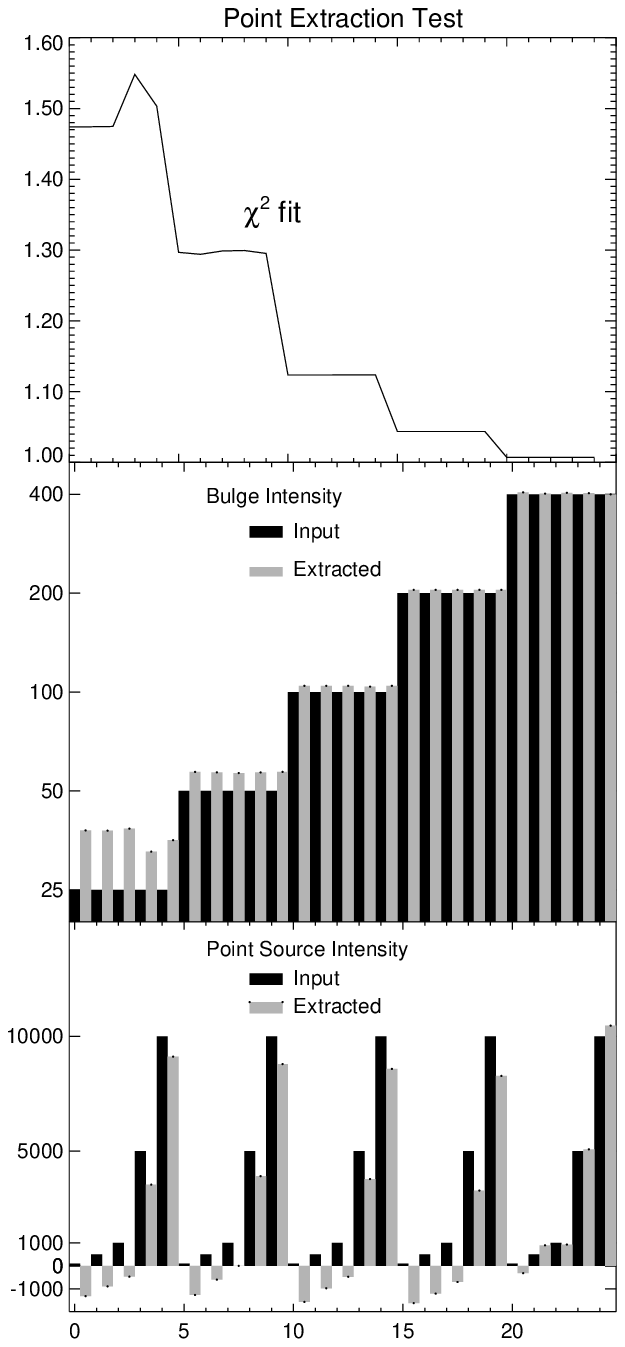}
\caption{Results for test of point extraction in 25 runs. The top
panel shows the $\chi_\nu^{\,2} $value. The bulge intensity increases after
every 5 runs resulting in a steep drop in $\chi^2$. The middle panel shows 25 input values
for bulge intensity and the corresponding recovered value adjacent to
each other. Panel 3 shows input point source intensities for each run
and the corresponding recovered value. See the text for an explanation.}
\label{fig:point}
\end{figure}

\subsection{Detecting a point source}
We aim to model a point source in addition to the disk and bulge.  The
\dev bulge intensity has a very sharp peak near the center, so it
should be easy for a point source to be swamped by the bulge unless it
is very bright.  The objective here is to find powerful sources, and
we will not be concerned with the point source if its intensity is
less than the bulge intensity over the central pixel.  We looked at
various strengths of the point relative to the bulge, by examining a
uniform grid of $ 5 \times 5 $ values of point and bulge intensities,
over which the bulge and point each varied by a factor of thirty.
There was no detection of a fictitious point. Weak points were
absorbed into the bulge, while strong points were extracted well, as
shown in Fig.~\ref{fig:point}.  For higher bulge intensities, the
point intensity was extracted better, because higher S/N far from the
center served to constrain the bulge intensity more precisely.

\section{Fitting real galaxy images}

The ultimate goal of this program is to extract parameters from a
large sample of galaxies of different morphological types. Work on a
large sample of spiral galaxies has been reported in RF96 and Byun,
1992.  We report here parameters we extracted for two ellipticals and
a lenticular galaxy.  These galaxies were chosen to illustrate the
algorithm's ability to find the disk and bulge self-consistently. We
selected the two ellipticals NGC 661 and NGC 1427 as simple tests, and
one disk galaxy, NGC 1381, to illustrate how well the program is
able to detect the disk.  We expect to extract similar scale lengths
in different bands and to get good fits ($\chi_\nu^2 < 2.0)$ in most
cases.

\subsection{Possible pitfalls}

Going from fitting models to fitting real galaxies has several
attendant dangers.  Significant errors in the PSF can cause the
detection of a fictitious point (Section 4.2.2). Since the PSF is
never known exactly, any extraction of a point must be examined very
carefully. Very close to the center, \dev law has not been shown to be
correct, and there is evidence from HST data to the contrary
(Byun~et~al 1996).  Without very precise knowledge of the PSF,
measuring systematic deviations from the law is not reasonable.
Images are often normalized, averaged, or background subtracted in
processing. Knowledge of the normalization used is required before we
get an accurate estimate of the S/N ratio which is a prerequisite for
determining the weighting function for our $\chi^2$ minimization.

\subsection{NGC 661}

\begin{figure}[p] 
\plotone{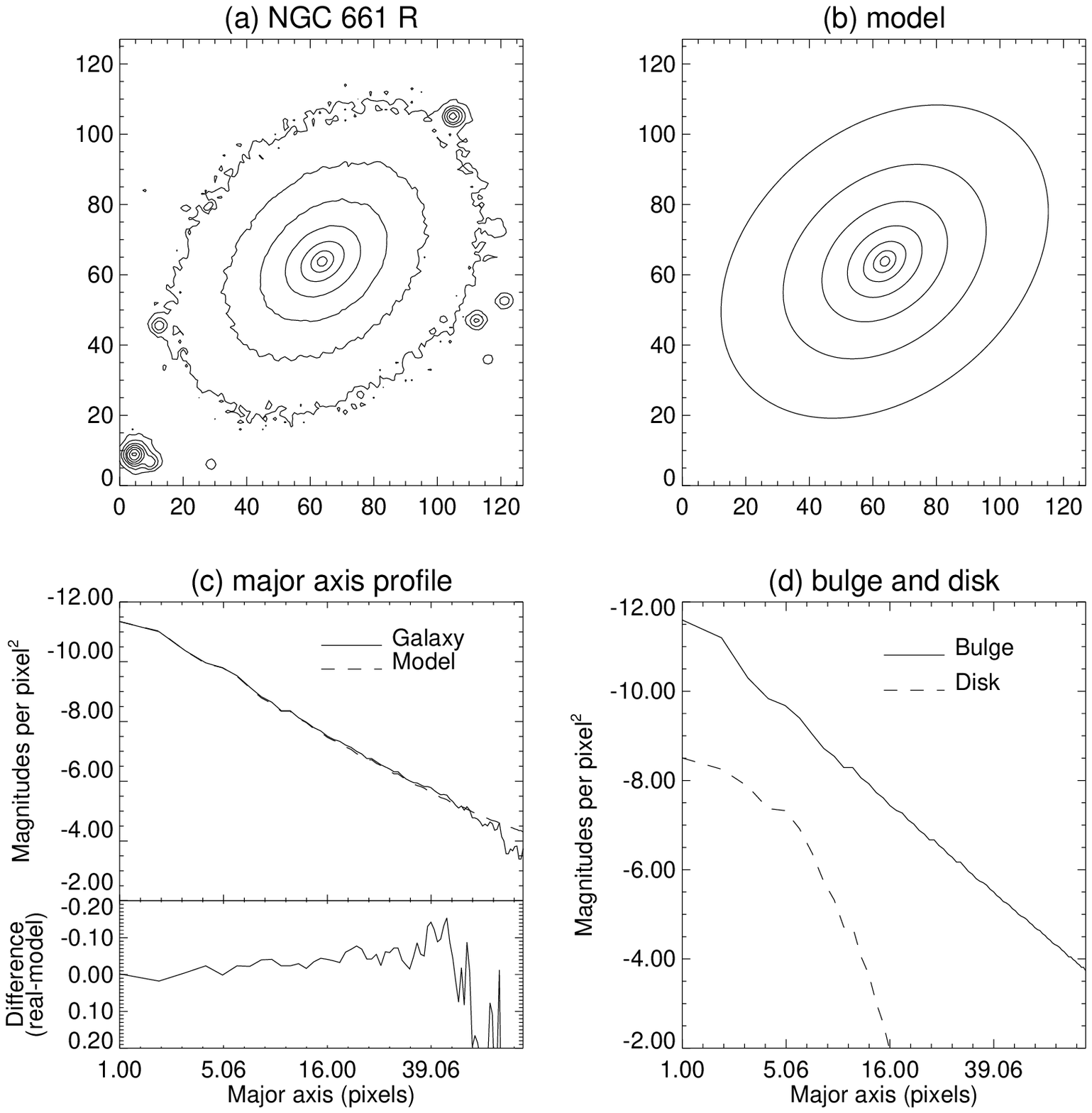}
\caption{NGC 661 in the R band and our fit with the center unmasked. (a)
is the 128 $\times$ 128 pixel image of the galaxy, with contours 
plotted at intervals of 
1 magnitude.  
(b) is the contour of our extracted model, with the same contour levels.
(c) is the
profile along the major axis.  When finished with the fitting, we step
along the major axis, and produce a list of galaxy intensities and
(PSF convolved) model, bulge and disk intensities at each point.  On 
this plot the model is represented by a dashed line, and the real
galaxy by a solid line. The difference between the real galaxy
magnitude and the model is plotted just below.  (d) shows the
decomposition into bulge and disk, also convolved with the PSF.  }
\label{fig:661R}
\end{figure}

NGC 661 has been classified as a possible cD galaxy with total magnitude in the B system $ m_{\rm B} = 12.88$ in the literature (RC3).  We used
images in two bands, V and R.  We ran the processing twice in each
band, once with the center masked out, and once with the center left
unmasked.  The results are shown in Table~\ref{tab:661R} and, for the
particular of the $R$ band analysis with the central pixels included,
in Figure~\ref{fig:661R}. It is clear from panel C of the figure that
the difference between the real and model galaxy profiles is very
difficult to see with the eye.  The difference becomes more
discernible at large radii because of the poorer signal to noise
ratio.  The observed and model profiles appear irregular because we
have plotted intensities along the major axis, rather than average
isophotal intensities.  The latter is usually done when considering
results from 1-D profile fits.
 
We have included a bulge and a disk in our fits to NGC\,661.  It is
seen from Table~\ref{tab:661R} that a weak disk, with total luminosity
$\sim4\%$ of the bulge, is detected.  The disk has a scale length
comparable to the FWHM of the PSF.  It is therefore not possible to
really distinguish between the disk and a point source. Moreover the
detected disk simply could be an artifact of departures of the bulge
profile, near the centre, from the de Vaucouleurs law form that we
have assumed in our analysis. The ambiguity is evident from the large
change in best fit disk scale length and intensity obtained in the $V$
band, in the cases with the central pixels masked and unmasked.
Inclusion of a disk in the fit can affect the bulge parameters
significantly as can be seen in the Table.  Since the disk parameters
cannot be trusted because of the reasons cited, the better alternative
for a galaxy like NGC\,661 is to fit a pure bulge profile.  We shown
the results of such a fit in Table~\ref{tab:661bulge}.  It is seen
that the bulge parameters obtained in the two filters are equal to
within the margin of error we found in parameter extraction for our
simulated images in Section~4.1.  The result obtained here is that the
bulge scale lengths in the two filters are equal within errors
expected in parameter extraction.  The results shown in
Table~\ref{tab:661R} are also consistent with this, but the bulge
scale lengths obtained there are $\sim25\%$ smaller.  This example
shows that features which are not clearly discernible given the PSF
and the signal to noise ratio, should not be included in profile fits.
Even though such features appear to be weak, they can significantly
alter the fitted parameter values of the stronger features.
  
\begin{table}[hbt]
\begin{center}
{\bf NGC 661: Bulge and Disk}\\
\medskip
\begin{tabular}{ |l|c|c|c|c|c|c|c|c|} \hline
& $\chi^2$  & $I_e$ &$r_e$ &$e_b$ &$I_d$ &$r_s$ &$e_d$& $D/B$ \\ \hline 
R  band& 1.097 &  540.3 &  21.11 & 0.3042 &   7283 &   2.23 &  0.755& 0.042 \\\hline 
R  band$^\dagger$&  1.075 &  522.6 &  21.45 & 0.3024 &   4681 &
2.701& 0.7133& 0.040 \\ \hline \hline
V band & 1.378 &  381.6&   19.63 & 0.3087 &   14078 &  0.0755 & 0.764
& 0 \\ \hline
V band$^\dagger$&  1.092 &  638.2 &  14.91 & 0.2945 &  3988 &  2.374
& 0.1104& 0.045 \\ \hline 
\end{tabular}
\caption{Extracted parameters for NGC 661 in the R and V bands. $\dagger$ :
Central pixels masked out.}
\end{center}
\label{tab:661R}
\end{table}
\begin{table}[hbt]
\begin{center}
{\bf NGC 661: Bulge only}\\
\medskip
\begin{tabular}{ |l|c|c|c|c|} \hline
& $\chi^2$  & $I_e$ &$r_e$ &$e_b$  \\ \hline 
R  band&  1.220 & 556.883 &  26.83 & 0.3315 \\ \hline
V  band&  1.220 & 584.746 &  26.90 & 0.3195 \\ \hline
\end{tabular}
\caption{Extracted parameters for NGC 661 in the R band :
Central pixels masked out}
\end{center}
\label{tab:661bulge}
\end{table}

\subsection{NGC 1381}

NGC 1381 is a galaxy of Hubble type SA0 and total  magnitude in the B
system $ m_{\rm
B} = 12.60 $ (RC3). We used images of this galaxy in the r,v, and g
Gunn filters.  Information about how this image was normalized was not
available to us, so our estimate of $ \chi_\nu^{\,2} $ should be taken
as a relative measure only. It has a significant, highly elliptical
disk, serving as a good test of our program's ability to extract
parameters when the isophotes are highly non-elliptical, and
one-dimensional fitting programs are inapplicable (BF95).

\begin{figure}[p]
\plotone{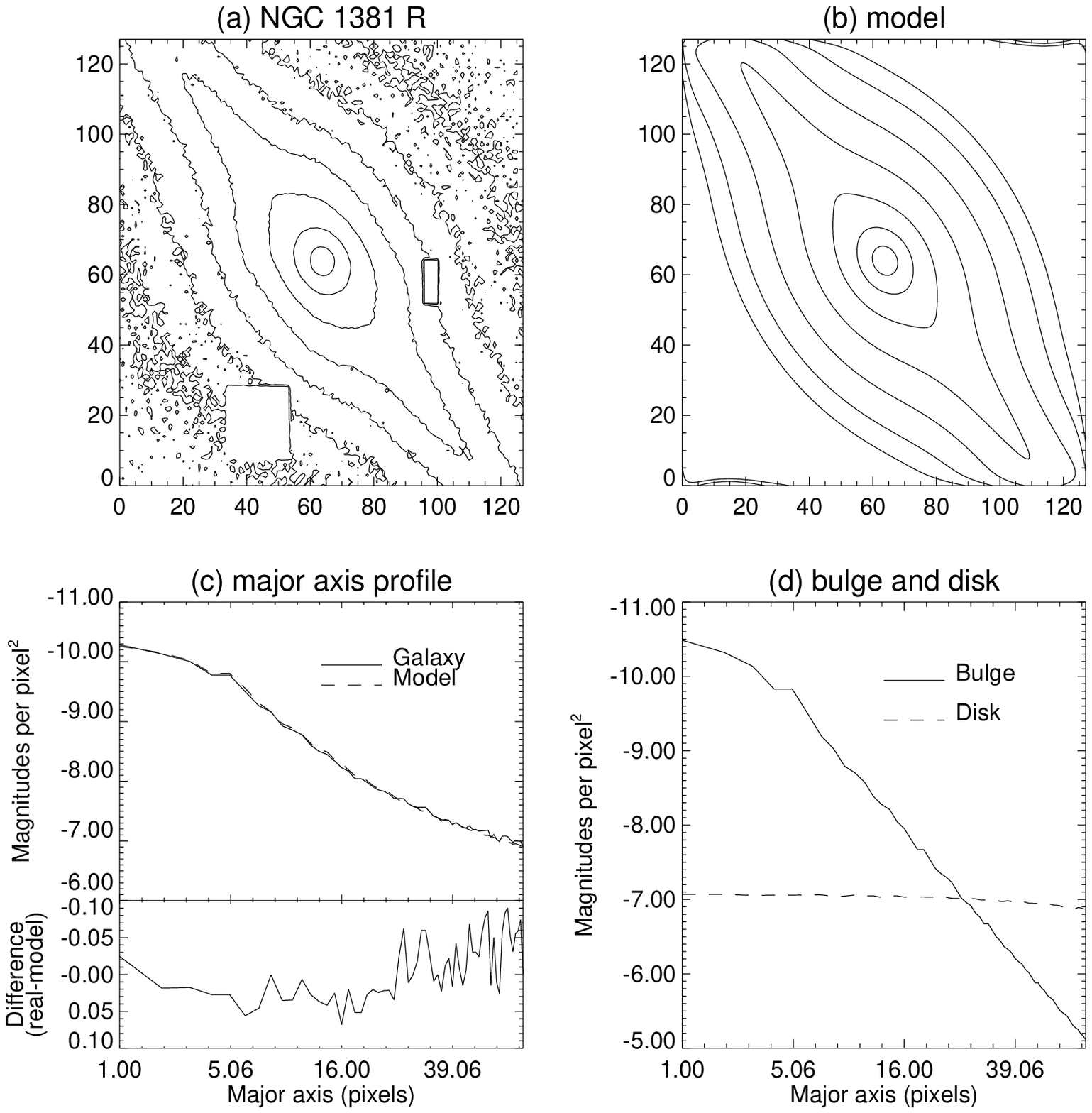}
\caption{NGC 1381 in the R band and our fit with the center
unmasked. For a description of the four panels see Fig.~\protect\ref{fig:661R}. The
rectangular regions in the image were blocked out in
preprocessing. Note that there are some deformed contours in the extreme upper right
and lower left of the model image.
These are an artifact of the convolution
in Fourier space, and were masked out in the fit. \protect\footnote{ The
convolution involves translating the unconvolved model and the
gaussian into Fourier space, multiplying them, and translating back
into the starting space. With $f$
representing the image, $g$ representing the gaussian, $h$
representing the convolved image, Fourier
transformation denoted by capitalization, (so $F$ is the Fourier
transform of f) and convolution by $\otimes$, $h = f \otimes g$ is
calculated  by first transforming $f$ and $g$ into $F$ and $G$, then
multiplying them, $H = F \times G $, and taking the inverse transform
to get $h$.  Since we are using a discrete transform, there are two
effects to consider-- the lowest frequency components of $F$ and $G$
will have a wavelength equal to the width of the image, and the
transform is the same as it would be if the image were endlessly
replicated on all sides. Then the convolution spreads out each points
intensity.  The pixels on the upper left and lower right of the
original image border on pixels that are zero initially, so with
convolution their values are lowered, and some intensity spills over
into the lower left and upper right.  For NGC 661 and NGC 1427, this
does not happen, because their values are very low near the edges.  If
it was important to avoid these edge effects, we could have padded the
image with zeros before convolving, but it was not important because
we masked out the affected areas in the fit.}}
\label{fig:1381R}
\end{figure}

Figure~\ref{fig:1381R} shows the R band decomposition of the this
galaxy. As in NGC 661, the difference between our model profile and
the galaxy profile is very difficult to see with the eye, and is never
more than 0.1 magnitude.

The bulge intensities are very similar both when the central pixels
are masked out and when they are unmasked. The bulge scale length
varies by $\sim 25\% $ in the v and r bands. The disk intensities are
recovered consistently as well.  We find that the disk scale length in
Table~\ref{tab:1381R} is very large compared to the size of the image,
making the disk intensity approximately constant along its length, as
is apparent from panel (d) of Fig~\ref{fig:1381R}.  Since the disk
ellipticity is high, we are looking at an edge on disk. This galaxy
illustrates the capability of 2-D fitting to detect components with
very high ellipticities. 1-D fitting that involves fitting ellipses to
the image to generate ellipse profiles.

Bulge and disk ellipticities are extracted very consistently,
with a maximum deviation of less than 4\%.
 
\begin{table}[hb] 
\begin{center}
{\bf NGC 1381}\\
\medskip
\begin{tabular}{|l|c||c|c|c|c|c|c|c|} \hline
& $\chi^2$  & $I_e$ &$r_e$ &$e_b$ &$I_d$ &$r_s$ &$e_d$&D/B \\ \hline 
Gunn g band &   1.75 &  122.1 &   41.1 & 0.2885 &  294.4 &  357.5 &
0.9767 & 51.08\\ \hline 
g band$^\dagger$&  1.419 &  117.4 &  42.72 & 0.2873 &  296.2 &  389.9
& 0.9775 & 58.85\\ \hline \hline
v band &1.378 &  31.78 &   32.4 & 0.2808 &  47.93 &    255 & 0.9654 & 26.16 \\ \hline 
v band$^\dagger$& 1.363 &  48.04 &  24.76 & 0.2821 &  47.21 &  195.8 &
0.9585& 22.28 \\ \hline \hline
r band & 2.887& 381.5 & 35.1& 0.2816& 828.2& 236.5& 0.9640 & 27.60 \\ \hline
r band$^\dagger$& 2.774 &  513.5 &  28.72 & 0.2584 &  867.8 &  174.6 &
0.9496& 17.49 \\ \hline \hline
\end{tabular}
\caption{Extracted parameters for NGC 1381 in the Gunn g, v, and r
bands. $\dagger$ : Central pixels masked out. }
\end{center}
\label{tab:1381R}
\end{table}

\subsection{NGC 1427}

This galaxy has been classified as a cD galaxy with total magnitude in
the B system $ m_{\rm B} = 11.85 $ (RC3). In this decomposition we
only found a bulge.  Since we had two images in the Gunn v and g
filters and once in the r filter, we performed the decomposition
separately on each image, to check the self-consistency of the
algorithm. We find that it agrees well with itself where we have two
images in the same wavelength, suggesting that the differences in
scale length between different filters may have physical
significance. The difference between scale lengths with and without
masking of the central pixels is remarkable in the R band image.  The
intensities compensate for the difference, however so that the final
$\chi_{\nu}^{\,2} $ are almost equal.  In the R band, the disk is not
wholly negligible unlike the other two bands where the $ D/B $ ratio
is extremely small.  This is, however, a small perturbation on the
bulge, as the bulge intensity is greater than that of the disk out to
a large radius (about 40 pixels), and it is doubtful, but possible,
that there is physical significance to this disk in the R band. The
extracted bulge ellipticity is almost constant in all filters. However
the extracted disk ellipticities show quite a wide scatter. This is
not surprising considering that the disk is very weak and our
estimate of the image PSF is only approximate.

\begin{table}[ht]
\begin{center}
{\bf NGC 1427}\\
\medskip
\begin{tabular}{|l||c|c|c|c|c|c|c|c|}	\hline
	    	     & $\chi^2$ & $I_e$ &$r_e$ &  $e_b$ &  $I_d$ &
$r_s$  &  $e_d$& D/B \\ \hline 
g band (1)  	     & 2.218 &  109.3 &  46.67 & 0.3008 &  -1516 &
0.5579 & 0.9963 & 0.00\\ \hline     
g band (1)$^\dagger$ & 1.974 & 83.39  &  56.97 & 0.2948 & 108149 &
1.2184 & 0.8906 & 0.16 \\ \hline 
g band (2)  	     & 1.660 & 169.1  &  46.15 & 0.2822 &   4204 &
1.9594 & 0.5199 & 0.01 \\ \hline 
g band (2)$^\dagger$ & 1.401 & 271.5  &  38.48 & 0.3193 &  -1976 &
7.6429 & 0.4728 & -0.01\\ \hline \hline
v band (1)  	     & 1.363 &  23.09 &  46.69 & 0.2819 &  -3775 &
0.451 & 0.1435 & 0.00\\ \hline 
v band (1)$^\dagger$ & 1.363 &  22.51 &  47.34 & 0.2784 &  1.448 &
175.9 & 0.9005 & 0.25 \\ \hline 
v band (2)  	     & 1.416 &  34.14 &  49.84 & 0.2750 & -49330 &
0.1935 & 0.0777 & 0.00 \\ \hline 
v band (2)$^\dagger$ & 1.413 &  32.58 &  51.64 & 0.2748 &  5.931 &
2465 & 0.7326 & 116.14\\ \hline \hline
r band   	     & 1.306 & 423.50 &  27.80 & 0.2502 &  495.8 &
43.865 & 0.4034 & 0.82\\ \hline 
r band$^\dagger$     & 1.334 & 187.68 &  54.68 & 0.2680 & 130.02 &
71.213 & 0.5827 & 0.33 \\ \hline 
\end{tabular}
\caption{Extracted parameters for NGC 1427 in the Gunn g, v, and r
bands.  Disk is much like either a point source or background in g and
v bands. Two images each in g and v and one image in the r band was
used. $\dagger$ : Central pixels masked out}
\end{center}
\end{table}
\begin{figure}[p]
\plotone{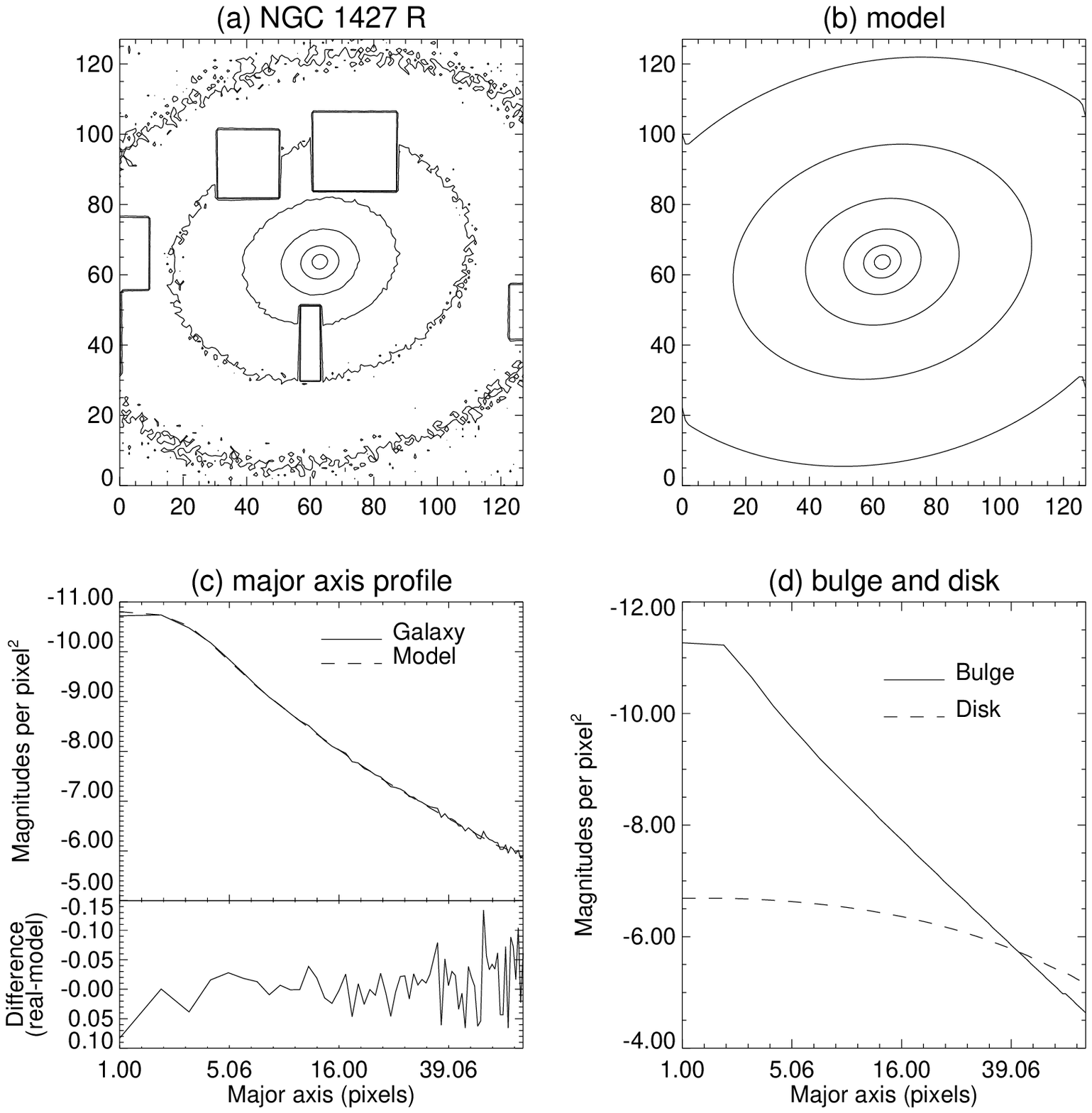}
\caption{NGC 1427 in the R band and our fit with the center
unmasked. For a description of the 4 panels see Fig.~\protect\ref{fig:661R}. }
\label{fig:1427R}
\end{figure}

\section{Error estimation}

In minimization problems, two  methods are commonly employed for parameter error estimation. The
first is to estimate the error from the second derivative of the function
being minimized with respect to the parameter under consideration. The
second is to gradually move away from the the minima until a predetermined
$ \chi^2 $ is exceeded. The second method will work for a single parameter
fit irrespective of whether the minima is parabolic in shape or of a more
complicated nature.  MINUIT can perform error estimations using both
methods.

In any multi--parameter minimization process, formal errors on the
parameters can be generated from the the covariance matrix of the fit only
if:
(i) the measurement errors are normally distributed
and,
(ii) the model is linear in its parameters
or
(ii) the sample size is large enough that the uncertainties in the fitted
parameters do not extend extend outside a region where the model could be
replaced by a suitable linearized model. It should be noted that this
criterion does not preclude the use of a non-linear fitting technique to
find the fitted parameters(Press et al, 1992).

Amongst the bulge and disk parameters that we use in the fit, two are
linear ($I_e $ and $ I_s $) and the rest are non-linear. Non-linearity is
highest for $ e_b $ and $ e_d $. Leaving all parameters free results in
rather large formal error bars on extracted parameters (about 20- 30 \%). The $
\chi^2 $ function is not parabolic near the minima causing erroneous error
estimation by MINUIT when the derivative method is used. Even moving away
from the minima till some $ \chi^2 $ is exceeded does not work as there
are multiple free parameters that are correlated. MINUIT is unable to
compute errors using this technique. Fixing the most non-linear
parameters i.e. the ellipticities to their extracted value enables MINUIT to compute formal
errors using this technique as the function can be approximated by a
linearized model. The errors are however still large. Fixing more
parameters reduces the error bars. Formal errors match those obtained from
parameter recovery in the 500 model test if only {\em one} parameter is
left free. Given the strong inherent non-linearity in our formulation of
the minimization problem, the problem of obtaining formal error bars on
extracted parameters, when more than one parameter is left free, may be
mathematically intractable.

\section{Conclusions}

Extensive tests on simulated galaxy images show that two
dimensional fitting is very successful at extracting the input
parameters in an overwhelming majority of cases, and that the cases
where it fails can be easily detected by looking at the $\chi^2$
value since  failure was always accompanied by a high
$\chi^2$ value.

One major limitation of our method is that it assumes that the
luminosity profile of the galaxy under consideration actually follows
the empirical laws we have chosen to model it irrespective of the
great variation seen in galaxy morphologies. Studies of the effects of
morphology such as dust 
absorption in disks modeled by Evans (1982)  on scale parameters are required if we are to
develop a reliable methodology to extract parameters for galaxies with strong
features such as bars, spiral arms etc. 

\subsection{Applicability of 2-D minimization methods}
Whether 1-D or 2-D methods are appropriate for fitting depends on
the morphology of the galaxy under consideration. In a very general sense
the following statements apply:

\begin{itemize}
\item For galaxies with very steep luminosity profiles 1-D fitting
provides a better solution than 2-D methods because in such galaxies,
a very large fraction of pixels have poor signal to noise which would
work against a good determination by the 2-D method.
\item When there is large isophotal twist in the galaxy, a 1-D method
works better than the 2-D one, because the 1-D method tends to follow
the twisting of the ellipses by changing the position angle with
radius while all  2-D methods proposed to date hold the position angle
constant.
\item When effects of shape parameters are significant then a 1-D
technique is better. As an example when a highly inclined disk is present, 1-D
fits might  miss it altogether.
\end{itemize} 

We intend to apply our technique to a sample of radio loud low $z$
elliptical galaxies currently being studied by us.

\acknowledgements
We thank S. G. Djorgovski for providing the data on NGC 1381 and NGC 1427 and
Dr. S K. Pandey for the data on NGC 661. We thank Ashish Mahabal and
S. K. Pandey for helpful comments and discussions.


\begin{thebibliography}
\bibitem{} Burnstein, D., 1979, \apj, 234, 435
\bibitem{} Byun and Freeman, 1995, \apj, 448, 563
\bibitem{} Byun et al, 1996, \aj, 111, 5
\bibitem{} Evans Rhodri, 1994, \mnras, 266, 511
\bibitem{} Freeman, K., 1970, \apj, 160, 811
\bibitem{} Kormendy, 1977, \apj, 217, 406 
\bibitem{} Knapen and van der Kruit, 1991, \aap, 248, 5
\bibitem{} Press et al, 1992, Numerical Recipes in C (Cambridge)  
\bibitem{} Schombert and Bothun, 1987, \aj, 93, 60
\bibitem{} Simien, 1989, in Le Monde des Galaxies, ed. H.G. Corwin,
Jr. \& L. Bottinelli (Berlin:Springer), 293
\bibitem{} de Jong R. S., 1996, \aaps, 118, 557
\bibitem{} de Vaucouleurs, 1948, Ann. d'Astrophys., 11, 247
\bibitem{} de Vaucouleurs et al, 1991, Third Reference Catalog of
Bright Galaxies (Berlin:Springer)
\bibitem{} van Albada, 1982, \mnras, 201, 939 
\end{thebibliography}
\end{document}